\documentclass[10pt]{article}
\usepackage{float}
\usepackage[symbol]{footmisc}
\usepackage{graphicx}
\usepackage[superscript,sort]{cite}
\usepackage{array}
\usepackage{bm}
\usepackage{appendix}
\newcolumntype{x}[1]{%
>{\centering\hspace{0pt}}p{#1}}%
\usepackage[normalem]{ulem} %strikeout a word
\usepackage[colorlinks=true,linkcolor=black,citecolor=black,urlcolor=black,bookmarks=true,breaklinks=true]{hyperref}
\usepackage[usenames]{color}
\usepackage{amsmath,amssymb}

\usepackage{amsthm,amscd,amsxtra,amsfonts,amsmath,amssymb,multirow}
\usepackage{wrapfig}
\usepackage[footnotesize]{caption}
\usepackage{subcaption}
\usepackage[tiny,compact]{titlesec}
\usepackage{threeparttable} %add footnote underneath the table
\usepackage{helvet}

\usepackage{booktabs}
\usepackage[table]{xcolor}
\usepackage{xcolor,colortbl}
\usepackage{placeins}
\usepackage{tikz}
\usetikzlibrary{shapes,arrows}
\usepackage[T1]{fontenc}
\usepackage[linesnumbered,ruled]{algorithm2e} %algorithm
\titlespacing*{\section}{0pt}{*0}{*0}
\titlespacing*{\subsection}{0pt}{*0}{*0}
\titlespacing*{\subsubsection}{0pt}{*0}{*0}
\titlespacing{\paragraph}{0pt}{*0}{*1}

\definecolor{MyPurple}{rgb}{1,0,1}

\newcommand{\beq}[1]{\begin{equation} \label{#1}}
\newcommand{\eeq}{\end{equation}}
\newcommand{\barray}{\begin{array}{ll}}
\newcommand{\earray}{\end{array}}

\usepackage{ulem}
\usepackage{xcolor,tikz}
\newcommand\reduline{%
	\bgroup\markoverwith
	{\textcolor{red}{\pgfsetfillopacity{0.2}\rule[-0.5ex]{2pt}{10pt}\pgfsetfillopacity{1}}%
		\textcolor{red}{\llap{\rule[0.4ex]{2pt}{0.4pt}}\llap{\rule[0.7ex]{2pt}{0.4pt}}}%
	}%
	\ULon
}
\newcommand\tealuline{%
	\bgroup\markoverwith
	{\textcolor{teal}{\pgfsetfillopacity{0.2}\rule[-0.5ex]{2pt}{10pt}\pgfsetfillopacity{1}}%
		\textcolor{teal}{\llap{\rule[0.4ex]{2pt}{0.4pt}}\llap{\rule[0.7ex]{2pt}{0.4pt}}}%
	}%
	\ULon
}

\usepackage{longtable} % longtable
\usepackage{threeparttable} %add footnote underneath the table
\usepackage{booktabs} %nice table
\begin{document}

\title{MLIMC: Machine learning-based implicit-solvent Monte Carlo  }

\author{Jiahui Chen$^{1}$, 
Weihua Geng$^{2}$\footnote{
		Corresponding author,		Email: wgeng@smu.edu },  ~  and  Guo-Wei Wei$^{1,3,}$\footnote{
		Corresponding author,		Email: wei@math.msu.edu}\\
$^1$ Department of Mathematics,
Michigan State University, MI 48824, USA.\\
$^2$ Department of Mathematics, 
 Southern Methodist University, Dallas, TX 75275, USA \\
$^3$ Department of Biochemistry and Molecular Biology,
Michigan State University, MI 48824, USA. \\ \\
\vspace{1cm}
This work is dedicated to Professor John ZH Zhang on the occasion of his 60th birthday
}

\date{\today}

%\centerline{ }
\maketitle

\begin{abstract}
Monte Carlo (MC) methods are important computational tools for molecular structure optimizations and predictions. 
When solvent effects are explicitly considered, MC methods  become very expensive due to the large degree of freedom associated with the water molecules and mobile ions. 
Alternatively implicit-solvent MC can largely reduce the computational cost 
by applying a mean field approximation to solvent effects and meanwhile
maintains the atomic detail of the target molecule. 
The two most popular implicit-solvent models are
the Poisson-Boltzmann (PB) model and the Generalized Born (GB) model 
in a way such that the GB model is an approximation to the PB model 
but is much faster in simulation time. 
In this work, we develop a machine learning-based implicit-solvent Monte Carlo (MLIMC) method by combining the advantages of both implicit solvent models in accuracy and efficiency. 
Specifically, the MLIMC method uses a fast and accurate PB-based machine learning (PBML) scheme to compute the electrostatic solvation free energy at each step.  
%PBML training data were generated using the matched interface and boundary Poisson Boltzmann (MIBPB) method, the most accurate PB solver for macromolecules.  
We validate our MLIMC method by using a benzene-water system and a protein-water system. 
%We present predicted total energy, root mean square deviation, errors and CPU time. 
%Compared with a few standard methods, 
We show that the proposed MLIMC method has great advantages 
    in speed and accuracy for molecular structure optimization and prediction. 
    
% {\bf Supplementary information:} Supplementary material is available at Bioinformatics online.

\end{abstract}


\newpage

\section{Introduction}\label{Introduction}
The determination of protein structures is of paramount importance for  structural biology and macromolecular study. However, not all protein structures can be determined with available experimental  techniques due to various limitations.  Computational methods offer important alternative approaches for structural determination and optimization \cite{wei2019}.  Indeed, molecular force field models and molecular dynamics \cite{Karplus2005,Alder1959, Rahman1964} can generate time-resolved trajectories of protein folding and protein-ligand binding predictions as well as structural ensemble simulations  \cite{Scheraga2007}.
In these simulations, mathematical models and numerical algorithms are imperative for achieving computational accuracy and efficiency. 
A large number of advanced algorithms have been developed to reduce the computational cost and improve the accuracy for  biomolecular simulations \cite{Sagui1999, Sutmann2011, DuanChen:2011a, Geng2013:3}.
A major difficulty of molecular dynamics is the long timescales associated with real molecular processes taking place in nature. Therefore, ignoring the requirement of having time-resolved trajectories of the molecular processes will immediately remove the difficulty. Indeed, it is sufficient for most studies to have a predicted representative ensemble of structures for a given process. This representative prediction can be generated by Monte Carlo sampling \cite{Frenkel2004}.

Monte Carlo method is one of the most of  popular approaches for biomolecular systems.  Under a physiological condition, biomolecules are immersed in and interact with surrounding water molecules and other possible co-factors. As such, Monte Carlo simulations of a biomolecule have to deal with a large number of solvent water molecules, which makes the simulations very expensive and sometimes, intractable. Additionally,  in Monte Carlo simulations, the biomolecular conformation is subject to random perturbations \cite{Metropolis1953}. These perturbations will inevitably result in the overlaps between the biomolecule and explicit solvent molecules, which leads to an unfavorable and non-representative structure. Implicit solvent models,  such as Poisson-Boltzmann (PB) \cite{Davis:1990a, Fogolari:2002}, polarizable continuum \cite{Tomasi:2005, Cossi:1996} and
Generalized Born (GB) methods \cite{Dominy:1999, Onufriev:2002, Tjong:2007b, Mongan:2007}
are developed to overcome this challenge by taking a mean field approximation of water molecules and resulting in a dielectric continuum. The GB method is faster than PB methods but it only provides an approximation for electrostatic energies. PB methods, derived from fundamental physical theories \cite{Beglov:1996, Onufriev:2000},  offer more accurate electrostatic analysis. PB model has been applied to the calculations of protein-protein and protein-ligand binding energies\cite{nguyen2017accurate},  the pH value predictions of protonation and/or deprotonation states of titration sites\cite{Jurrus2018}, and drug design \cite{wang2019end}.
To seek for an accurate, efficient, and robust numerical solver,
a large number of numerical methods have been developed for the PB model, including
finite difference method (FDM) \cite{Jo:2008},
finite element method (FEM) \cite{Baker:2001b},
and boundary element method (BEM) \cite{Geng2013:3, LuBenzhuo:2013}.
Among this variety of numerical explorations,
the FDM has the most enfranchisement such as Amber PBSA \cite{PBSA:2008},
Delphi \cite{lidelphi:2012}, 
APBS \cite{Baker:2001b, Jurrus2018}, 
MIBPB \cite{zhou2008highly, Geng:2007a, DuanChen:2011a,DDNguyen:2017a,geng2017two}, and 
CHARMM PBEQ \cite{Jo:2008}. Among them, MIBPB is the solely available second-order accurate method and has been used to calibrate the GB method in Amber \cite{forouzesh2017grid}, where PB methods are generally very expensive. In addition, the molecular surface involved in all the aforementioned method with corresponding software developed, such as ESES\cite{liu2017eses}, Nanoshaper\cite{decherchi2013general}, and MSMS\cite{sanner1996reduced}.

Over the past a few years,  machine learning, including deep learning, has had tremendous success in science and engineering. 
Especially, convolutional neural networks have proved their ability to automatically extract features and recognize patterns from relatively simple but large datasets. Deep learning has a growing dominance in important applications such as handwriting recognition, speech recognition, and drug discovery \cite{hughes2015modeling, lusci:2013, nguyen2019mathematical}.
Aided by the availability  of  quality databases, new algorithms,  graphics processing unit (GPU), and high-performance computers, various machine learning approaches have established in many classical computational problems such as solvation free energies, protein-ligand binding affinities, mutation impacts, toxicity, partition coefficients, protein B-factors,  etc \cite{korotcov2017comparison,karimi2018deepaffinity,jimenez2018k,wang2017improving, DDNguyen:2017d, ZXCang:2017b,ZXCang:2018a,KDWu:2018a, KDWu:2018b,ching2018opportunities}. Additionally, deep learning neural networks are also applied in computational protein design \cite{wang2018computational}, stability changes of protein induced by mutations \cite{cao2019deepddg,cang2017analysis}, and calculations of protein energy\cite{wang2020combining,chen2018neural}.

Recently, we developed a Poisson-Boltzmann based machine learning (PBML) model, which can compute the solvation free energy of macromolecules in the solvent  with the GB speed and the PB accuracy \cite{Chen}.
We assume that all of the macromolecular electrostatic solvation free energies follow a probability distribution, which can be sampled by the PB model.  Our idea is based on a representability hypothesis and a learning hypothesis. The representability hypothesis states that the solvation free energy of a molecule can be described by the features of atom interactions and their geometric relations in the solvent. Thus, we can construct feature vectors to characterize the molecular electrostatic distribution. In our learning hypothesis, we assume that a machine learning model can be trained based on training labels and corresponding features for a sufficiently large training set of molecules. Additionally, advanced machine learning algorithms can give accurate predictions of the electrostatic potential for a new molecule which has the same probability distribution with the training set. In our approach,  training labels are computed from MIBPB and features are generated using multiscale weighted colored subgraphs \cite{DDNguyen:2017d}. 
   
In the present work, we apply our newly developed PBML model to compute molecular solvation free energies in the implicit-solvent Monte Carlo simulations, which typically require millions of samplings. The new machine learning-based implicit-solvent Monte Carlo model can guarantee the accuracy of the implicit-solvent Monte Carlo model while dramatically speeding up existing implicit-solvent Monte Carlo algorithms. 

This manuscript is organized as follows. Section~\ref{sec_BCT} gives a brief introduction of molecular force fields, Monte Carlo methods, and implicit solvent models. The PBML model is introduced in this section as well, which includes the Poisson-Boltzmann equation, Generalized Born model, and multiscale weighted colored subgraphs.  Section~\ref{sec_result} presents the results of structural predictions of benzene and {the human hyperplastic discs protein (PDB: 1i2t)}~\cite{deo2001x} in water. We demonstrate that the PBML model is more accurate and faster than commonly used PB solvers and thus, can significantly reduce the computational time of implicit-solvent Monte Carlo simulations. A summary is given in Section~\ref{sec_conclusion}.\\

\section{Methods and algorithms} \label{sec_BCT}

In this section, we briefly review biomolecular force fields, the  Monte Carlo methods, and implicit solvent models, followed by  the Poisson-Boltzmann based machine learning model.\\

\subsection{Biomolecular force fields}
The quality of molecular simulations depends crucially on molecular force fields to offer a physical representation of molecular interactions and energy distributions.  Molecular force fields typically describe molecular interactions in terms of classical molecular mechanics of atoms. The potential energies of atomic interactions are approximated by a set of mathematical functions, modeling the bonded and non-bonded components.   These functions consist of a set of free coefficients, which are obtained by approximating either the results of elaborate quantum mechanical calculations, or experimental data. One of the advantages of biomolecular force field approach is its computational efficiency.
The potential energy can be efficiently computed  at the molecular level comparing to other methods, such as quantum mechanical approaches, which deal with electrons  
\cite{CaseD.A.;CheathamT.E.;DardenT.;GohlkeH.;LuoR.;MerzK.M.;Onufriev2005, 
Lindorff-LarsenK.;PianaS.;PalmoK.;MaragakisP.;KlepeisJ.L.;DrorR.O.;Shaw2010}.
Additionally, the forces in molecular dynamics can be evaluated analytically from molecular force fields.

A variety of molecular force fields have been developed for various purpose. In this work, we adopt the popular and simple Amber ff99SB force field 
\cite{Lindorff-LarsenK.;PianaS.;PalmoK.;MaragakisP.;KlepeisJ.L.;DrorR.O.;Shaw2010}.
The Amber force field for governing the potential energy consists of the following terms,
\begin{equation}
E = \sum_{\rm bonds}k_b(r-r_0)^2 + \sum_{\rm angles}k_{\theta} (\theta-\theta_0)^2
+ \sum_{\rm dihedrals}V_n[1+\cos(n\phi-\gamma)] +
 \sum_{i=1}^{N-1}\sum_{j=i+1}^{N}\Bigg[\frac{A_{ij}}{R_{ij}^{12}} - \frac{B_{ij}}{R_{ij}^{6}}
  + \frac{q_iq_j}{\epsilon_1 R_{ij}} \Bigg]
\end{equation}
where $k_b$, $k_{\theta}$, and $V_n$ are force constants. Here, $r$, $\theta$, and $\phi$ are bond length, angle and dihedral angle with  $r_0$, $\theta_0$, and $\gamma$ being optimal bond length, optimal angle and proper dihedral angle, respectively. 
The first three terms in the energy expression describe the bonded energy of the molecular system.
The last term represents the Lennard-Jones interactions and electrostatic interactions, where $N$ is the number of atoms in the molecular system,
$R_{ij}$ is the distance between $i$th and $j$th atoms,
$A_{ij}$ and $B_{ij}$ are Lennard-Jones parameters, 
$q_i$ is the atom charge, and $\epsilon_1$ is the dielectric constant.\\

\subsection{Monte Carlo methods}
In this session, we provide a brief introduction of the molecular dynamics and the Monte Carlo method. We start from statistical mechanics and show that the calculation of the physical property of a solute-solvent system using molecular dynamics is computationally expensive or even intractable  \cite{Frenkel2004}.
Then, we introduce  Metropolis's Monte Carlo method for biomolecular simulations \cite{Metropolis1953}.

The classical expression for the partition function $Q$ of a solute-solvent system is
\begin{equation}
\label{eq_partition_fcn}
Q=c\int d\textbf{r}d\textbf{p} \exp[-\mathcal{H}(\textbf{r},\textbf{p})/k_BT],
\end{equation}
where $\textbf{r}=\{\textbf{X},\textbf{Y}\}$ stands for the atomic coordinates of a solute $\textbf{X}$ and solvent $\textbf{Y}$,
$\textbf{p}$ stands for the corresponding momenta,
$c$ is a physical constant as specified below, $k_B$ is the Boltzmann constant 
and $T$ is the temperature of the system.
The function $\mathcal{H}(\textbf{r}, \textbf{p})$ is the Hamiltonian of the system.
It describes the total energy
of an individual system as summation of the kinetic energy $\mathcal{K}$ and the potential energy $E$:  $\mathcal{H}=\mathcal{K}+E$, 
where $\mathcal{K}$ is a quadratic function of the momenta.
For a system of $N$ identical atoms, one has $c=1/(h^{3N}N!)$ using the Planck constant $h$.    
Under the assumption that all of the other physical observables $A$ of interest depend only on the positions, i.e., $A=A(\textbf{r})$, the integration over the momenta can be carried out analytically in a classical mechanical treatment. As a result, the expected value of a physical observable of interest is given by
\begin{equation}
\label{eq_thermodynamic}
\langle A \rangle = \frac{\int d\textbf{r} A(\textbf{r}) \exp[-\beta E(\textbf{r})]}{\int d\textbf{r} \exp[-\beta E(\textbf{r})]},
\end{equation}
where $\beta = 1/k_BT$. 
Evaluating $\langle A \rangle$ requires numerical techniques, such as 
%Simpson's rule or Trapezoidal rule 
quadrature rules for the integration. Since each particle moves in a three dimensional (3D) space, the total number of degrees of freedom is $3N$ for a system of $N$ atoms.
If each dimension is integrated with a mesh size of $m$ points, the total number of points for the integration is $m^{3N}$, which is computationally prohibitive. 
%Thus even for a small number of $m$ in a rough approximation, 
%$m^{3N}$ can be a huge number 
%because an average protein typically has about 5000 atoms.

The complexity in evaluating Eq.~(\ref{eq_thermodynamic}) can be significantly reduced by using the Monte Carlo sampling.
Indeed,  Metropolis \textit{et al.} \cite{Metropolis1953} suggested an efficient Monte Carlo scheme to approximate the ratio in Eq.~(\ref{eq_thermodynamic}). Let us denote the probability density function in finding a microstate in the canonical ensemble in a configuration $\textbf{r}$ by
\begin{equation}
\label{eq_prob_den}
P(\textbf{r})=\frac{\exp[-\beta E(\textbf{r})]}{\int d\textbf{r} \exp[-\beta E(\textbf{r})]}.
\end{equation}
According to this probability function, we can perturb randomly selected points in the configuration. Hence, the number of points $n_i$ generated per unit volume in the neighborhood of $\textbf{r}$
is equal to $N_{mc}\times P(\textbf{r})$ for the average of $A(\textbf{r})$, which is
\begin{equation}
\label{eq_MC_approx}
\langle A \rangle \approx \frac{1}{N_{mc}} \sum_{i=1}^{N_{mc}}n_iA(\textbf{r}_i),
\end{equation}
{where $N_{mc}$ is the total number running in Monte Carlo simulations.}
Equation~(\ref{eq_MC_approx}) shows that all states of ensemble contribute to the average equally.
Therefore, Metropolis Monte Carlo method starts at a given configuration
$\textbf{r}_0=\{\textbf{X}_0, \textbf{Y}_0\}$ and
next perturbs the configuration by a defined transformation
with a new configuration $\textbf{r}_1=\{\textbf{X}_1, \textbf{Y}_1\}$.
The probability to accept the new configuration is
\begin{equation}
\label{eq_paccept}
p_{acc} = \min(1, \exp(-\beta(E(\textbf{r}_0)-E(\textbf{r}_1)))).
\end{equation} 
If the new configuration is rejected, the previous configuration is retained and the method repeats another random perturbation. This process iterates until the iteration number equals to a fixed number.
It is shown that the structure in the system will approach
the Boltzmann distribution, if the perturbations satisfy the condition
\begin{equation}
\label{eq_per_condition}
\pi(\textbf{r}_i)p_{ij}=\pi(\textbf{r}_j)p_{ji},
\end{equation}
where $\pi(\textbf{r}_i)$ is the probability of the system in configuration $\textbf{r}_i$
and $p_{ij}$ is the probability to perturb the configuration 
from state $\textbf{r}_i$ to state $\textbf{r}_j$ \cite{Metropolis1953}.\\

\subsection{Implicit solvent models}
Implicit solvent models are class of  multiscale techniques for reducing the dimensionality of a solvent-solute system. They retain the crucial electrostatic interactions between a biomolecule and its solvent environment without modeling solvent molecules explicitly. A variety of two-scale implicit solvent models  has been developed, such as  the Poisson-Boltzmann (PB) model \cite{Fogolari:2002} and the generalized Born (GB) model \cite{Dominy:1999, Onufriev:2002, Tjong:2007b, Mongan:2007}. One desirable application of implicit solvent models is the Monte Carlo simulations of biomolecule in solvent, which is relatively easy to implement.
The basic derivation for molecular implicit solvent models relies on statistical mechanics. For more detail, the reader is referred  to the literature \cite{Roux1999}.
Essentially, the molecular  solvation free energy can be given by 
\begin{equation}
\label{eq_sol_free_energy}
\Delta G_{\rm solv} =  \Delta G_{\rm elec} + \Delta G_{\rm nonpol},
%\nabla G_{\rm solv} =  \nabla G_{\rm elec} + \nabla G_{\rm nonpol},
\end{equation}
where $\Delta G_{\rm elec}$ represents the electrostatic contribution of the solvent-solute interaction, 
and $\Delta G_{\rm nonpol}$ denotes the nonpolar energy in the reversible work needed to insert  a fixed configuration molecule into the solvent with all solute charges set to zero.
Here $\Delta G_{\rm nonpol}$  is proportional to  the solvent accessible surface area. The molecular  solvation free energy is used in our implicit-solvent Monte Carlo method to represent solvent-solute interactions. \\

\subsection{ Poisson-Boltzmann based machine learning (PBML) model}

In this section, we briefly discuss the Poisson-Boltzmann based machine learning (PBML) model
\cite{Chen}, which is applied   to compute $\Delta G_{\rm elec}$ in Eq.~(\ref{eq_sol_free_energy}). 
Our PBML model involves three major components, 
i.e., training labels, molecular features, and learning algorithms.  
Our training labels for a large training set of molecules are generated from solving the Poisson-Boltzmann (PB) equation. Our molecular features for both the training set and the test set  constitute two parts, 
a GB part and a correction part. The latter is computed from multiscale weighted colored subgraphs \cite{Chen}.  

\paragraph{The Poisson-Boltzmann (PB) model}
The PB model considers the solute biomolecule with $N_c$ fixed charges as the interior domain $\Omega_1$, and the solvent, including free ions, as the exterior domain $\Omega_2$.
The interface $\Gamma$ separates these two  domains. The PB model is given as
\begin{equation}
-\nabla \cdot \epsilon({\bf r}) \nabla \phi({\bf r}) + \bar{\kappa}^2({\bf r})\phi({\bf r}) =
\sum_{k=1}^{N_c} q_k \delta({\bf r}-{\bf r}_k),
\label{eqNPBE}
\end{equation}
For ${\bf r} \in \mathbb{R}^3$, $\phi(\bf{r})$ is the electrostatic potential, $\epsilon(\bf{r})$  dielectric constant given by 
\begin{equation}
\epsilon(\bf{r})=\left\{
\begin{array}{ll}
\epsilon_1, \quad {\bf{r}} \in \Omega_1,
\\
\epsilon_2, \quad {\bf{r}} \in \Omega_2,
\end{array}
\right.
\label{eqEpsilon}
\end{equation}
In the PB model, $\bar{\kappa}$ is the screening parameter with the relation $\bar{\kappa}^2 = \epsilon_2 \kappa^2$ where $\kappa$ is the inverse Debye length measuring the ionic effective length. To ensure the continuity of electrostatic potential and flux density across the interface $\Gamma$, the PB equation is associated with following interface conditions
\begin{equation}
\phi_1({\bf r}) = \phi_2({\bf r}),
\quad
\epsilon_1 \frac{\partial\phi_1({\bf r})}{\partial {\bf n}}=\epsilon_2 \frac{\partial \phi_2 ({\bf r})}{\partial {\bf n}},
\quad
{\bf r} \in \Gamma
\label{eqInterface}
\end{equation}
where $\phi_1$ and $\phi_2$ are electrostatic potential from the solute domain $\Omega_1$ and the solvent domain $\Omega_2$, and $\bf{n}$ is the outward unit normal vector on $\Gamma$.

The solvation free energy can be obtained from the PB model by 
\begin{equation}
\Delta G_{\rm elec}^{\rm PB}= \frac{1}{2}\sum_{k=1}^{N_c}q_k (\phi(\bf{r}_k) -\phi_0(\bf{r}_k) )
\label{solvationEnergy}
\end{equation}
where $\phi_0(\bf{r}_k)$ is the free space solution to the PB equation assuming no solvent-solute interface.
To solve the PB equation,  we apply the accurate and robust 2nd order MIBPB solver \cite{DuanChen:2011a, DDNguyen:2017a} developed in our group, which applies rigorous treatment on geometric complexity, interface condition, and charge singularity. The $\Delta G_{\rm elec}^{\rm PB}$ results generated by MIBPB solver for a set of macromolecules are used as the training labels in the representability hypothesis.

\paragraph{The Generalized Born (GB) model}
Having described the labels for our machine learning training,  we  discuss the molecular feature construction for both machine learning training and test, which involves the GB model. 
As a fast approximation to the PB model, the GB model compute the electrostatic solvation free energy by 
\begin{equation}
\label{GB_eqn}
\Delta G^{\rm GB}_{\rm elec} \approx \sum_{i,j}\Delta G^{\rm GB}_{ij} = -\frac{1}{2} \Big( \frac{1}{\epsilon_1}-\frac{1}{\epsilon_2} \Big) \frac{1}{1+\alpha\beta} \sum_{i,j} q_i q_j \Big( \frac{1}{f_{ij}(r_{ij}, R_{i}, R_j)} + \frac{\alpha\beta}{B} \Big),
\end{equation}
where $R_i$ is the effective Born radius for $i$-th atom, $r_{ij}$ is the distance between atoms $i$ and $j$, $\beta = \epsilon_1/\epsilon_2$, $\alpha = 0.571412$, and $B$ is the electrostatic size of the molecule.
The function $f_{ij}$ is given as
\begin{equation}
\label{f_ij}
f_{ij} = \sqrt{r^2_{ij}+R_iR_j {\rm exp}\Big( -\frac{r^2_{ij}}{4R_iR_j} \Big)}.
\end{equation}
The effective Born radii $R_i$ is calculated by the following boundary integral
\begin{equation}
\label{eqn:Born_radii}
R^{-1}_i = {\Big( -\frac{1}{4\pi} \oint_{\Gamma } \frac{{\bf r}-{\bf r}_i}{|{\bf r}-{\bf r}_i|^6} \cdot \text{d}\bf{S} \Big)}^{1/3}. 
\end{equation}
In Eq.~(\ref{eqn:Born_radii}), the MSMS package \cite{MSMS} is used to generate  the triangulation discretization of the molecular surface for the numerical surface integral on $\Gamma$.

\paragraph{Multiscale weighted colored subgraphs}
The weighted colored subgraph (WCS) use the notion $G(V, E)$ with vertices $V$ and edges $E$ to describe the atomic interactions in a protein of $N$ atoms. 
The vertices is defined as
\begin{align}
V=\{({\bf r}_i,\alpha_i)| {\bf r}_i\in \mathbb{R}^3, \alpha_i\in \mathcal{C}, i=1,2,\dots, N\},
\end{align}
where $\mathcal{C}=\{{\rm  C, N, O, S, H} \}$ contains all the commonly occurring element types in a protein. Each vertex is an atom labeled by both its position ${\bf r}_i$ element type $\alpha_i$, for $i=1,\cdots N$. 

%According to different biomolecular systems, we need to change the set  $\mathcal{C}$.
The edge $E$ relates the pairwise interactions, 
which are defined as a colored set $\mathcal{P}=\{\alpha \beta\}$ with $\alpha, \beta \in \mathcal{C}$.
For $\mathcal{C}$ defined above, 
$\mathcal{P}=\{{\rm CC, CN,CO,CS,CH, NN,NO,NS,NH,OO,OS,OH,SS,SH,HH}\}$
and
we defined the partition of $\mathcal{P}$ as $\mathcal{P}_k$, $k=1,2,...,15$
such that $\mathcal{P}_1=\{{\rm CC}\}$, $\mathcal{P}_2=\{{\rm CN}\}$ and so on.
The set of involved vertices $V_{{\cal P}_k}$ is a subset of $V$
containing all atoms involved in forming the pair in $\mathcal{P}_k$.
For instance, $\mathcal{P}_2=\{{\rm CN}\}$
contains all carbon-nitrogen atom pairs
and $V_{{\cal P}_2}$ contains all carbon and nitrogen atom vertices 
in the protein. Based on these configuration, 
all the edges for pairwise atomic interactions in the WCS description 
are defined by
\begin{align}
E_{\mathcal{P}_k}^{\sigma,\tau,\zeta} = \{ \Phi^\sigma_{\tau,\zeta}(\|{\bf r}_i - {\bf r}_j\|)~|~\alpha_i\beta_j \in\mathcal{P}_k; i=1,2,\dots,N_{\alpha}, j=1,2,\dots,N_{\beta}\},
\end{align}
where $\|{\bf r}_i - {\bf r}_j\|$ defines the Euclidean distance between $i^{th}$ and $j^{th}$ atoms,
$N_{\alpha}$ and $N_{\beta}$ are numbers of type $\alpha$ and $\beta$ atoms,
$\sigma$ indicates the type of radial basic functions (e.g., $\sigma=\text{L}$ for Lorentz kernel, $\sigma=\text{E}$ for exponential kernel), $\tau$ is a scale distance factor between two atoms and $\zeta$ is a parameter of power in the kernel (i.e., $\zeta=\kappa$ for $\sigma=\text{E}$, $\zeta=\nu$ for $\sigma=\text{L}$).
In this model, we use generalized exponential functions
\begin{align}
\Phi^{\rm E}_{\tau,\kappa} = e^{-(\|\mathbf{r}_i-\mathbf{r}_j\|/\tau(r_i+r_j))^{\kappa}},\quad \kappa>0,
\end{align}
and generalized Lorentz functions
\begin{align}
\Phi^{\rm L}_{\tau,\nu}(\|\mathbf{r}_i-\mathbf{r}_j\|)= \dfrac{1}{1+(\|\mathbf{r}_i-\mathbf{r}_j\|/\tau(r_i+r_j))^{\nu}},\quad \nu>0,
\end{align}
where $r_i$ and $r_j$ are, respectively, the van der Waals radius of the $i^{th}$ and $j^{th}$ atoms.
Finally, the features for describe the electrostatics interactions and geometric properties are expressed as
\begin{equation}
\mu^{k,\sigma,\tau,\nu,w}=\sum_{i=1}^{N_{\alpha}}\sum_{j=1}^{N_{\beta}} w_{ij}\Phi^{\sigma}_{\tau,\nu}(\|\mathbf{r}_i-\mathbf{r}_j\|),
\quad \alpha_i\beta_j \in \mathcal{P}_k,
\end{equation}
where  $w_{ij}$ is a weight function assigned to each atomic pair with
$w_{ij}=1$ for atomic rigidity or 
$w_{ij}=q_j$ for atomic charge.
Since we have 15 options of the colored subsets $\mathcal{P}_k$,
we can obtain corresponding 15 subgraph centralities $\mu^{k,\sigma,\tau,\nu,w}$, for $k=1,2,\dots,15$. 
By varying kernel parameters $(\sigma,\tau,\nu,w)$, 
one can achieve multiscale centralities for multiscale weighted colored subgraph (MWCS) \cite{bramer2018multiscale},
which can be the features.
 
With labels and features described above, we can construct the machine learning model to predict the solvation free energy of new macromolecules. Specifically,  using MIBPB results as labels, and GB and MWCS results as features, we train gradient boosting decision trees (GBDTs) for the solvation free energy prediction.\\

\section{Results} \label{sec_result}

In this section, we demonstrate the performance of the proposed MLIMC method numerically. 
{First, we describe the Poisson-Boltzmann based machine learning (PBML) model 
for computing protein electrostatic solvation energies, followed by the illustration of the accuracy and efficiency of the model. The use of the PBML model for electrostatic interactions in the MC simulations is introduced.} Our main idea is to replace  time-consuming electrostatic calculations by using  our PBML model. The efficiency of our new MLIMC model is also examined. 
Finally, we  validate the proposed MLIMC method by two cases. 
Case 1 is a small molecule, benzene, with initial atom position randomly protruded. Our MLIMC method is used to reconstruct the benzene molecule in solvent.  
Case 2 is  a relatively larger molecule, protein {(PDB: 1i2t)} with 61 amino acid residues. In this case, 
we stretch the last two residues of 1i2t using steered molecular dynamics 
and then we try to restore the equilibrium configuration by using the proposed MLIMC method.  Both simulations are carried out at a temperature of $27^{\circ}\text{C}$, the dielectric constants are $\epsilon_1=1$ in the molecule and $\epsilon_2=80$ in the solvent, the MSMS \cite{MSMS} mesh density is set as  2, and the Debye-Huckel constant is set as $\kappa = 0.1257$\AA$^{-1}$. There are three kernels used to generated features for machine learning, which are $(\text{E}, 0.3, 2, 1)$, $(\text{E}, 4.7, 2, q_j)$, and $(\text{L}, 4.2, 5, 1)$.

To measure the performance,  we use the root-mean-square deviation (RMSD) of atomic positions in length units (\AA), defined as
\begin{equation}
\label{eq_rmsd}
\text{RMSD}(\textbf{v}, \textbf{w}) = \sqrt{\frac{1}{N}\sum_{i=1}^{N}\Big( (v_{ix}-w_{ix})^2 + (v_{iy}-w_{iy})^2 + (v_{iz}-w_{iz})^2 \Big)},
\end{equation}
where $\textbf{v}, \textbf{w} \in \mathbb{R}^{N \times 3}$ are vectors of positions of the $N$ atoms at two different MC samplings.
Moreover, we also present relative errors of the total energy 
measured by comparing the energy for a MC sampling $E_{MC}$, 
and the energy for the  equilibrium state  $E_{SS}$ as
\begin{equation}
\label{eq_relerr}
\text{e}_\text{ttl} = \frac{|E_{SS}-E_{MC}|}{|E_{SS}|}\times 100\%.
\end{equation}
We compute the RMSD and errors between Monte Carlo sampling results and the original molecular structure for every $100$ Monte Carlo steps for both cases. The core code was written in C/C++ and a cython wrapper calling the core code for performing adds-on functions and applications. Our simulations are produced on a desktop with an i5 7500 CPU and 16GB memory.\\

\subsection{{PBML model }}
{The MLPB model used in Monte Carlo simulation is a pre-trained model. The training set includes 3706 protein structures from the PDBbind v2015 refined set\cite{PDBBind:2015}. This refined set was selected from a general set of 14,620 protein-ligand complexes. A data pre-processing (i.e. adding force field parameters) is required before a PB solver can be used for electrostatics calculations. Though the PDBbind refined set consists of protein-ligand complexes, only protein structures are applied for calculations. These protein structures are adjusted by the protein preparation wizard utility of the Schrodinger 2015-2 Suite \cite{schrodinger} with default parameters unless filling the missing side chains is required.}

{The training set covers a wide range of proteins in different sizes with atom numbers from 997 to 27,713. 
The current training set can be expanded to an even larger group of proteins.  
However, from our test, we conclude that expanding training set will not significantly improve the trained model, 
thus the size of the current training set is sufficiently large.}

{The purpose of PBML is to implement a machine learning predictor of PB electrostatic solvation free energies for various proteins efficiently and accurately without explicitly solving the PB equation. 
Gradient boosting decision tree method is selected for this supervised learning task
because of its efficiency. 
To accuracy of the PBML model is maintained by the accurate electrostatic free energy of solvation as the label
calculated by the  MIBPB solver. Once a trained PBML model is obtained, the MIBPB solver will not be called anymore. 
Using the learned PBML model only requires calculating features on the prediction of electrostatic solvation free energies for new compounds, which is rapid. }
%Lastly, the parameters of GBDT model are learning rate 0.05, the number of estimators 5, the minimum number of samples required to be at a leaf node 1, and the minimum number of samples required to split an internal node 2.}

\subsection{Efficiency of the PBML model}
\begin{figure}[h]
	\centering
	\begin{subfigure}[b]{0.3\textwidth}
%\centering
\includegraphics[height=2.5in]{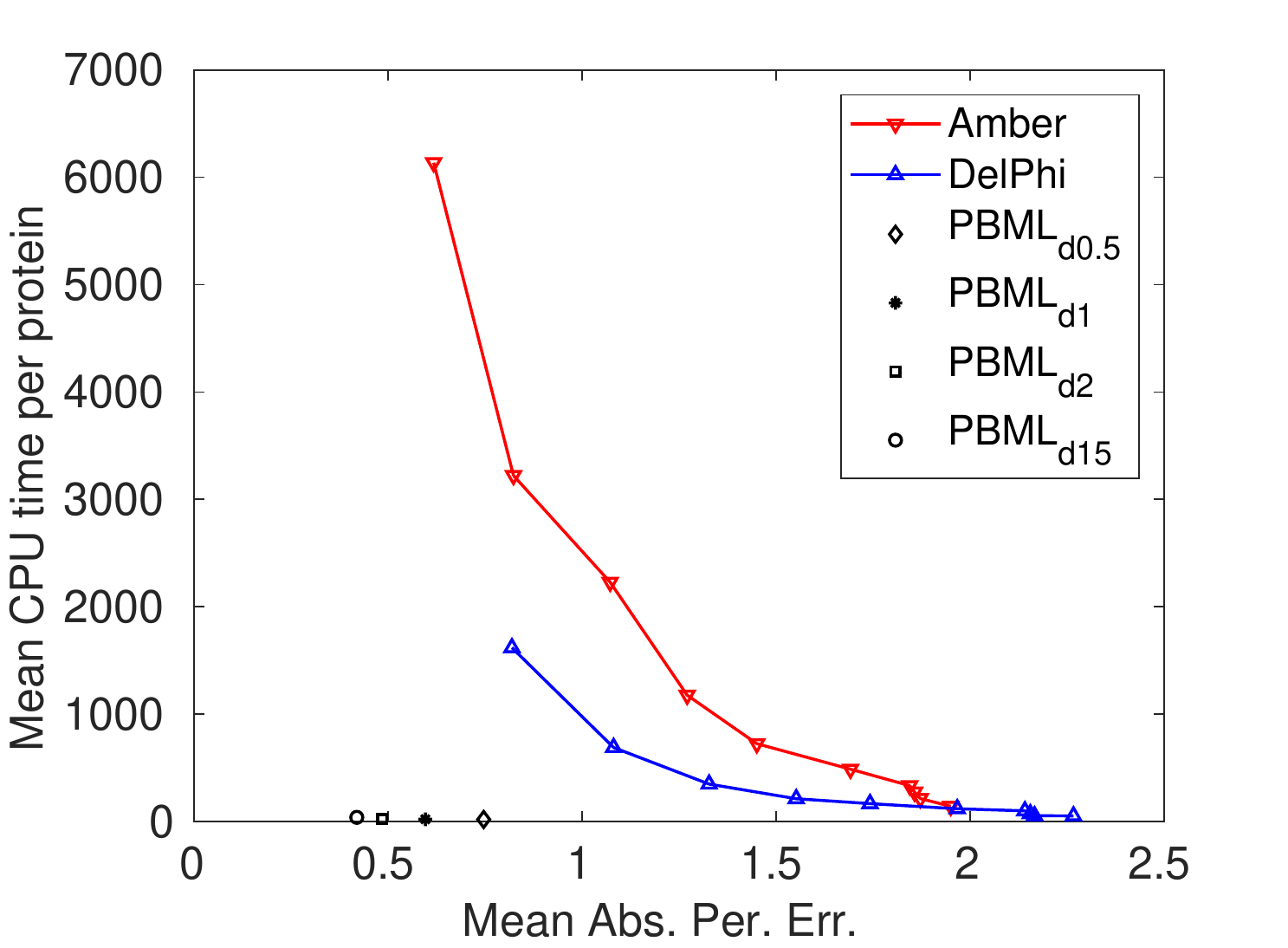}
%\caption{3owj}
\end{subfigure}
\hfill
\begin{subfigure}[b]{0.5\textwidth}
%\centering
\includegraphics[height=2.5in]{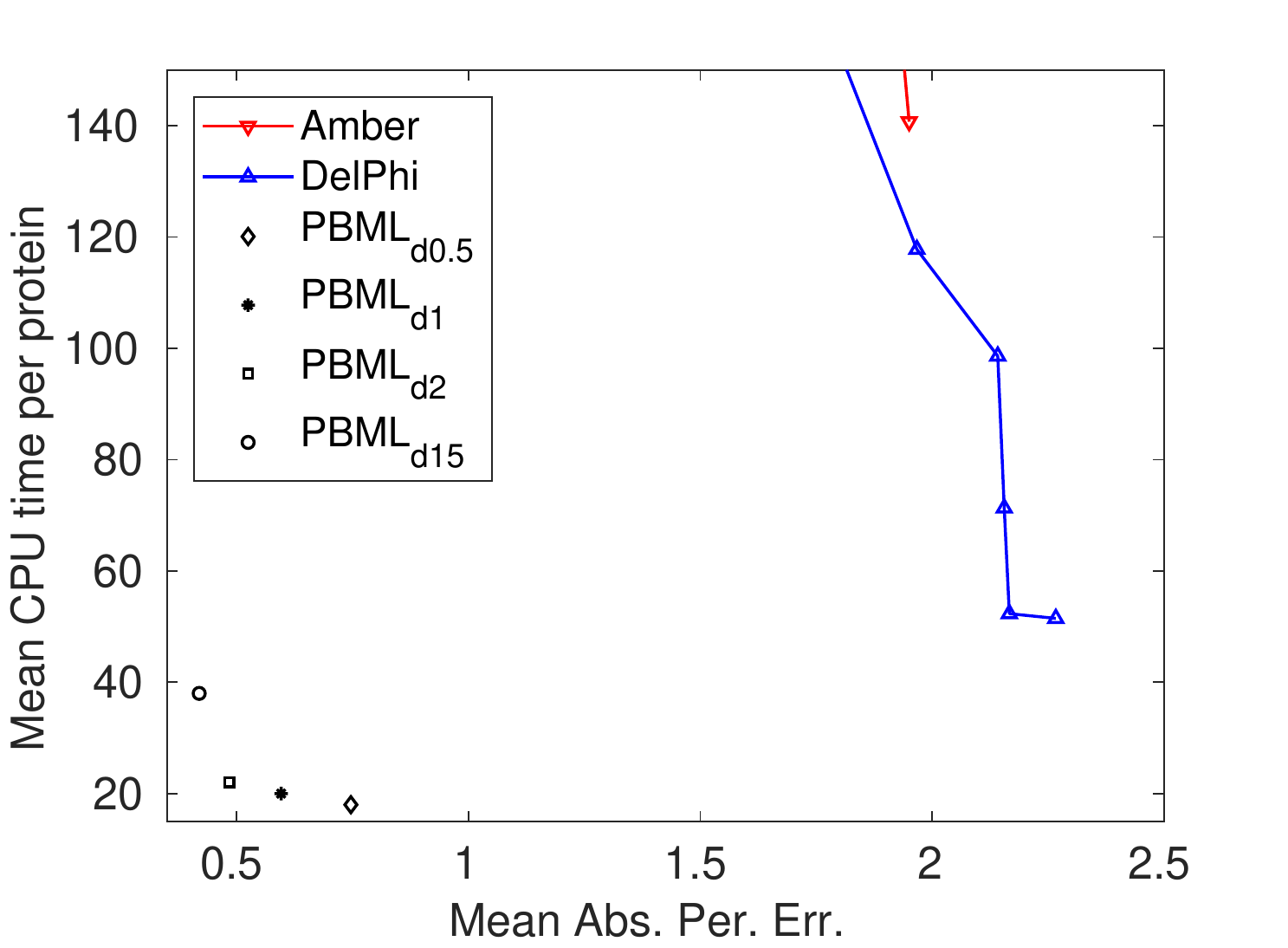}
\end{subfigure}
\caption{
Comparison of the mean CPU time (second) per protein and the mean absolute percentage errors of Amber, DelPhi and machine learning predictions of the electrostatic solvation free energies using the test set  of 195 proteins. 
		Left: Results of 	Amber and  DelPhi were obtained  at ten different mesh sizes from 0.2 to 1.1 \AA; Results of PBML were obtained at four MSMS densities (number of vertices per \AA$^2$) at 15, 2, 1, and 0.5. 
		Right: A zoom-in plot of  the left plot for small CPU time.  
	}
	\label{fgr:Time}
\end{figure}

Figure~\ref{fgr:Time} shows the results for computing solvation energy 
on 195 proteins from PDBbind v2015 core set \cite{PDBBind:2015} 
using PBML, Amber, and Dephi. The results are shown 
in terms of the average CPU time per protein versus the mean absolute percentage errors.  From Fig.~\ref{fgr:Time}(a), we can see PBML is more accurate and much faster than standard PB solvers such as DelPhi and Amber PB. Figure~\ref{fgr:Time}(b)
gives more details by zooming into the region where CPU time is small to distinct the CPU time used by the PBML using different MSMS density.   

%determines the total CPU time of the whole simulation.
We here add a few notes about how we improve the PBML model in addition to machine learning.  
We notice that in the energy and feature calculations,  
every term has a degree of freedom associated with the number of atoms,
except the computation of the effective Born radii $R_i$ in Eq.~(\ref{eqn:Born_radii}),
which depends on the number of surface triangles $M$.
Since  $M\gg N$, faster evaluating of Eq.~(\ref{eqn:Born_radii})
can significantly accelerate the entire Monte Carlo process. 
In our present implementation,
instead of taking the integral in Eq~(\ref{eqn:Born_radii}) on each triangle,
we take the integral on a neighborhood of each vertex. This treatment nearly doubled the efficiency of the GB method 
since number of vertices is about half of number of triangles on the surface.  
In addition, applying a cut-off can also further improve the GB method.  

\subsection{{MLIMC model}}

{The assembling of MLIMC includes the implementation of   empirical potential energy functions (except electrostatics) and the prediction of electrostatics for each step on Monte Carlo simulations. 
The conformation of the target protein is perturbed randomly on each step. 
The new conformation is directly accepted if it shows a lower energy or is accepted with a probability 
determined by the Boltzmann distribution if it shows a higher energy.}
{As the MLPB model is pre-trained before simulations, the Monte Carlo simulation does not include the time for solving the PB equation,  resulting in  much reduced time for MLIMC simulations.}

\subsection{Efficiency of the MLIMC model}
We   show that the high efficiency of the MLPB model will significantly improve the efficiency of the MLIMC model. 
\begin{table}[htp]
\small
\caption{Average CPU time for one step MLIMC simulation using Amber, DelPhi and PBML for electrostatic solvation free energy on the 195 protein dataset.   
Results of Amber and DelPhi were obtained at $0.2\text{\AA}$ and $0.5\text{\AA}$ mesh sizes, 
and that from PBML uses mesh density $2$. 
The average CPU time includes all computations needed for Monte Carlo evaluations. 
The PB error is obtained relative to the electrostatic solvation energy computed from MIBPB solver
with grid size $h=0.2\text{\AA}$.
}
\begin{center}
\renewcommand{\arraystretch}{1.3}
\begin{tabular}{c|c|c|c|c|c }
\hline
&&\multicolumn{2}{c|}{$h=0.2\text{\AA}$}&\multicolumn{2}{c}{$h=0.5\text{\AA}$}\\\hline
PB solvers & PBML & Amber & DelPhi & Amber & DelPhi \\ \hline
Avg. CPU time (sec) & 25 & 6136 & 1621 & 1177 & 214\\ \hline
PB error (\%) & 0.484 & 0.618 & 0.819 & 1.271 & 1.552 \\
\hline 
\end{tabular}
\end{center}
\label{tb:cputime}
\end{table}%
Table~\ref{tb:cputime} shows the mean CPU time of one Monte Carlo step
and the mean absolute percentage errors of Amber, DelPhi and
PBML predictions of the electrostatic solvation free energies of the 195 proteins.
The mean CPU time for each protein includes the computations for the total energies, in which computing electrostatic is the dominant component.
%The results of Amber and DelPhi were obtained at $0.2\text{\AA}$ and $0.5\text{\AA}$ mesh sizes.
%For the PBML, the MSMS density is 2 when generating above results.
Clearly, the machine learning method has the highest accuracy but lowest CPU time.
For the same accuracy level ($<1\%$), the estimation of the mean CPU time for a one-million-step Monte Carlo simulation is $6.136\times10^{8}$s, $1.621\times10^{8}$s and $2.5\times10^{6}$s  for using Amber, DelPhi and PBML,  respectively.
Even with compromised accuracy for DelPhi and Amber at gird size of $0.5\text{\AA}$,  the MLIMC with PBML will be $47$ times faster than that with Amber and $8$ times faster than that with DelPhi.
Next we show some MC simulation results using MLIMC on the benzene molecule and {the human hyperplastic discs protein (PDB:1i2t)}.

\subsection{Test case one: Benzene molecule}

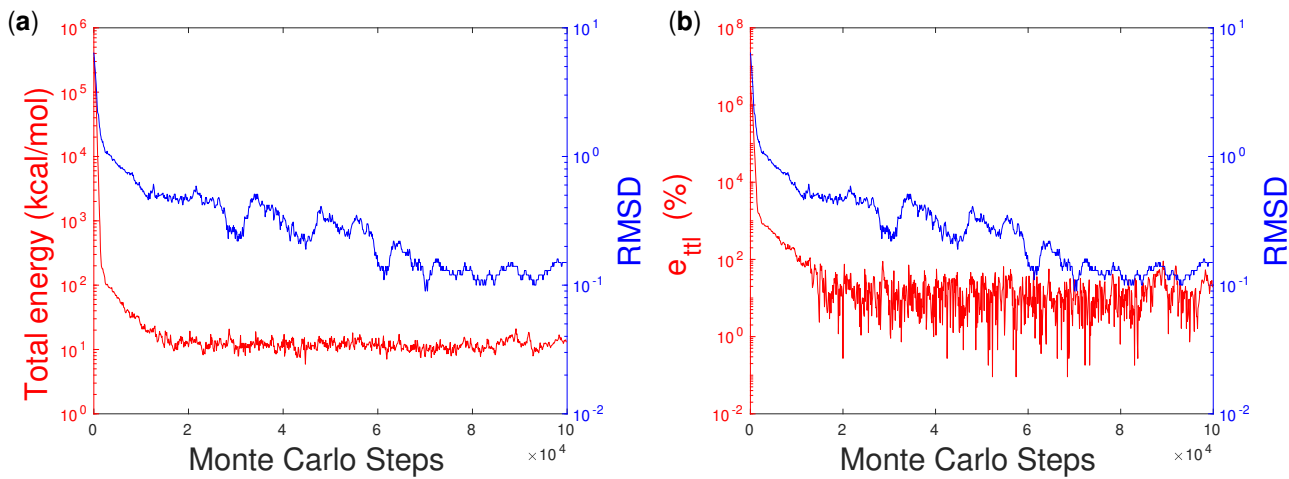
\begin{figure}
\centering
\includegraphics[height=2.55in]{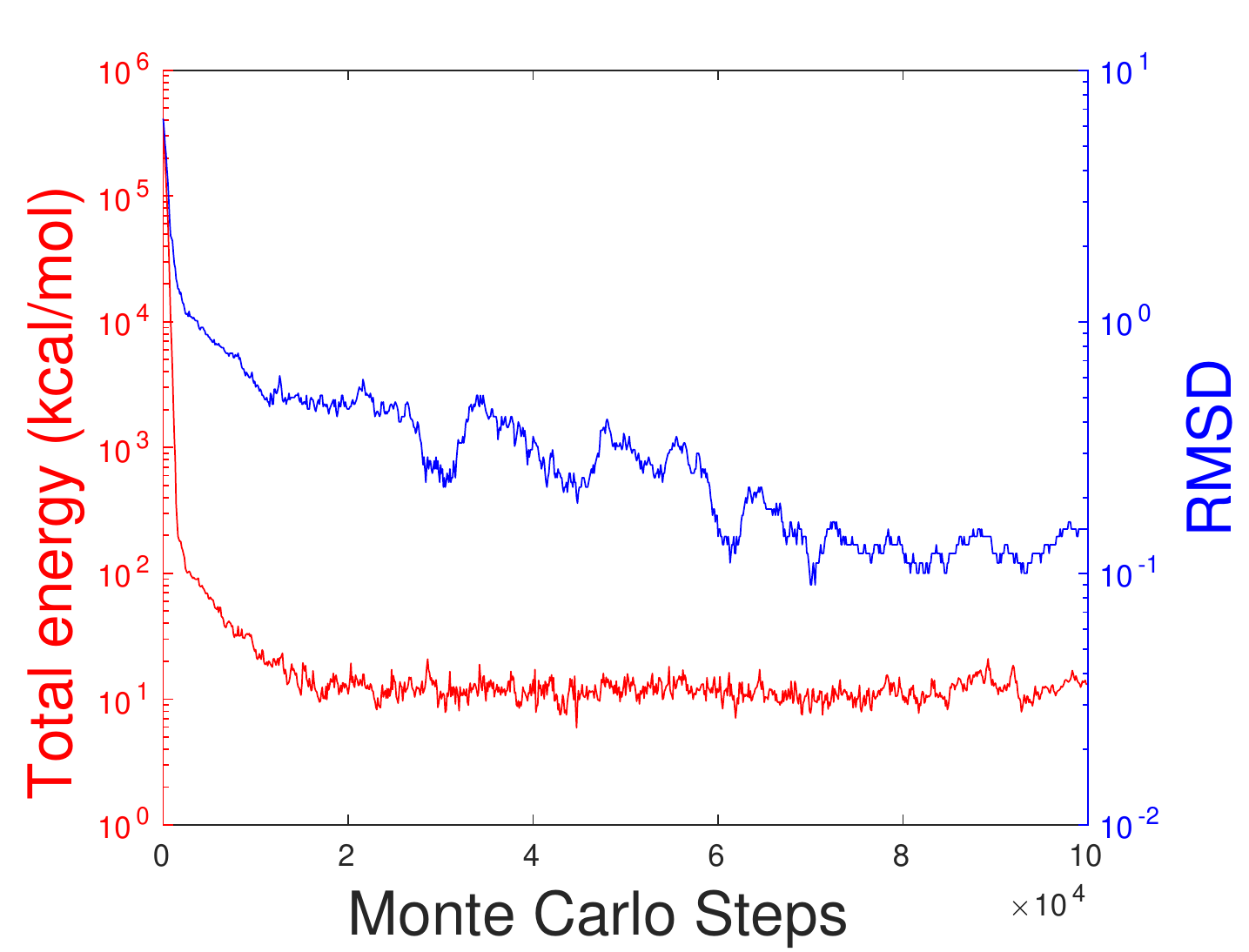}
\includegraphics[height=2.55in]{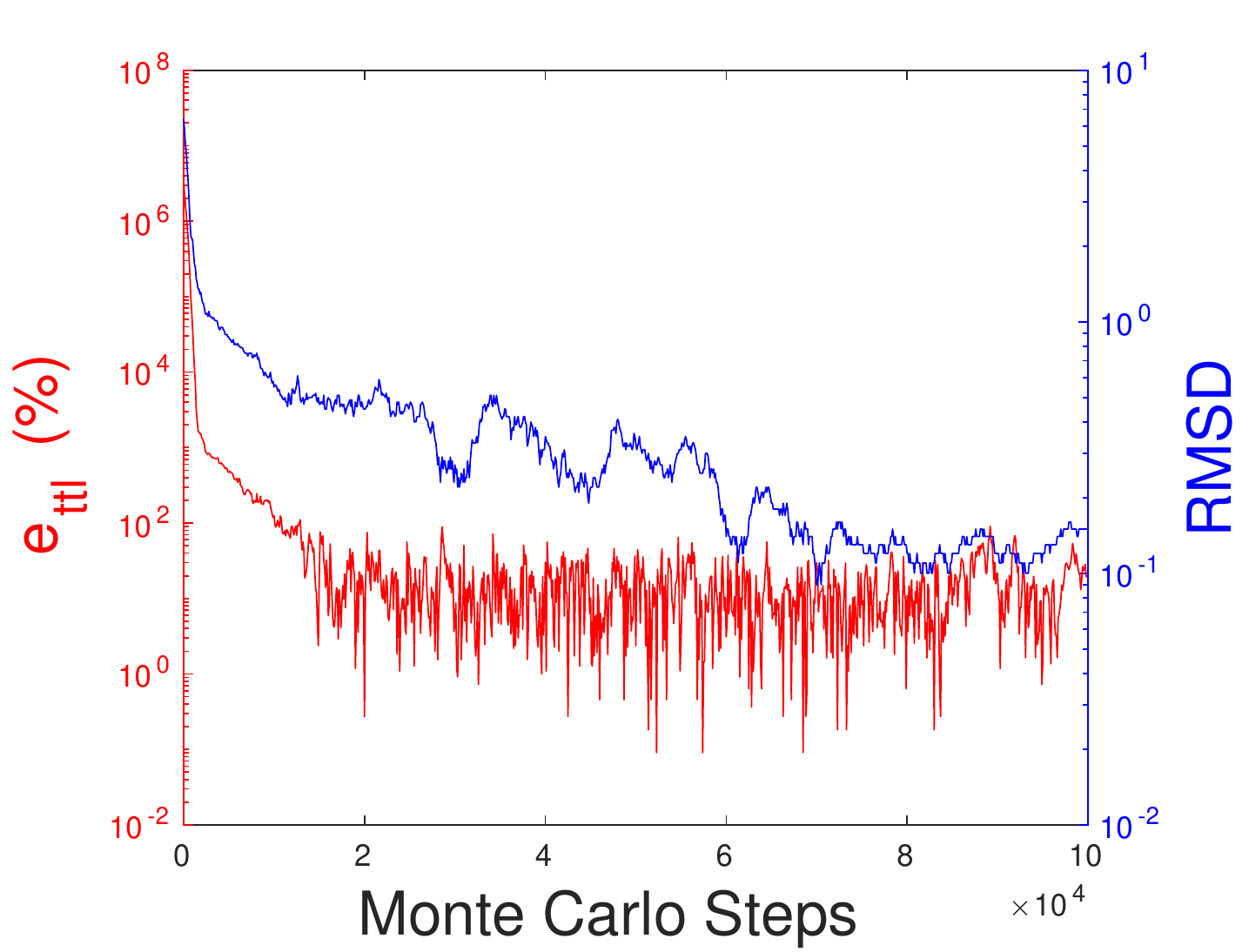}
\begin{picture}(0,0)
\put( -490, 170){(\textbf{a})}
\put( -240, 170){(\textbf{b})}
\end{picture}
\caption{MLIMC simulation of benzene in solvent.
(a) the red curve is the total energy calculated by our implicit-solvent Monte Carlo model and  
the blue curve is the root mean square deviation of the atomic positions {on each Monte Carlo step to the non-protruded  one}. 
(b) the red curve is the error of total energies {e$_\text{ttl}$ defined by Eq.~\ref{eq_relerr}} and the blue curve is the same RMSD {as the} left figure.}
\label{fgr:bnz}
\end{figure}

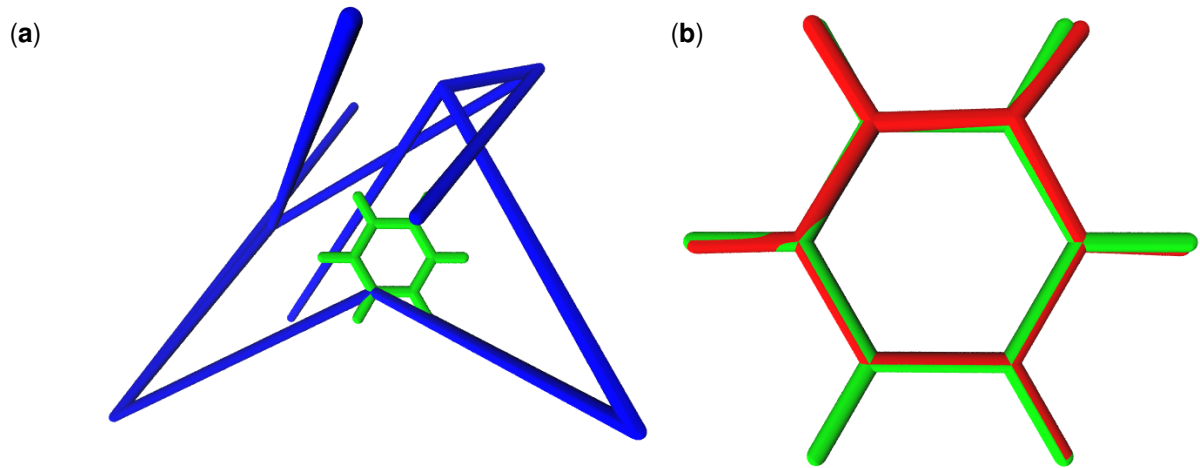
\begin{figure}
\centering
\includegraphics[height=2.6in]{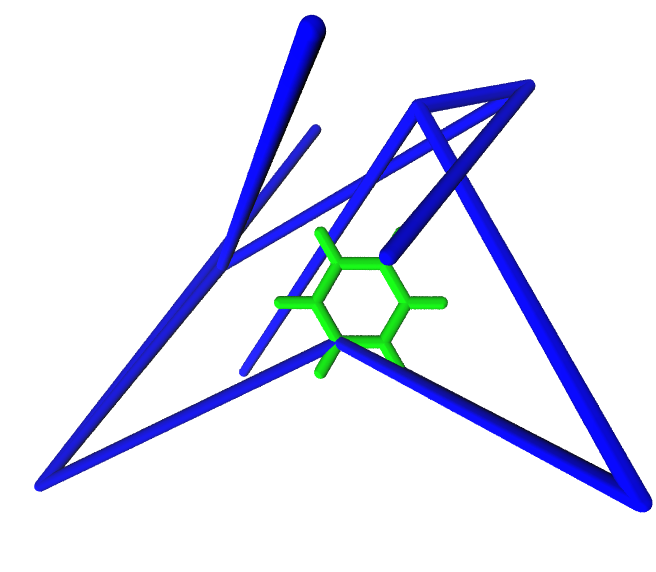}
\includegraphics[height=2.6in]{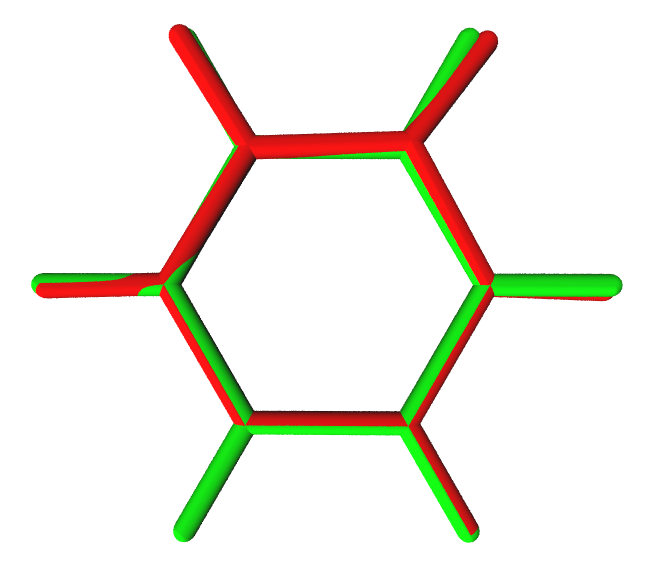}
\begin{picture}(0,0)
\put( -463, 170){(\textbf{a})}
\put( -213, 170){(\textbf{b})}
\end{picture}
\caption{Illustration of MLIMC simulations of a benzene molecular in solvent. (a) the blue structure is the randomly perturbed atom positions and the green one is the benzene structure in steady state. (b) the red one is the benzene structure after MLIMC sampling compared with the equilibrium structure in green. Pictures are produced with VMD  \cite{VMD}. }
\label{fgr:bnz_vmd}
\end{figure}

Our first case is a Benzene molecule with some atomic position randomly perturbed. 
In detail, we fixed three atoms at equilibrium positions in order to 
have the prediction and the comparison structure in the same plane, and perturb the coordinates of the remained nine atoms in $(\rho, \theta, \phi)$ directions by uniformly distributed random numbers in $([0, 10], [0, 2\pi], [0,\pi])$. The initial RMSD is $6.42 \text{\AA}$ as compared with the equilibrium position.
We will try to perform a MC simulation on this perturbed molecule to see if the original steady status can be obtained.  
Figure~\ref{fgr:bnz}(a) shows the total energy and RMSD vs MC steps, 
from which we can see that the total energy of benzene in solvent starts at $349123.61$ kcal/mol
and converges to the range of $(5, 15)$ kcal/mol after the first $20,000$ MC steps. It stays in a convergent range for the rest MC steps.
The RMSD initially is $6.42 \text{\AA }$ 
and ends around $0.15\text{\AA}$.
It decreases rapidly 
as the total energy for the first $20,000$ steps.
After $20,000$ steps, the total energy converges with only slightly oscillation,
and the RMSD keeps the decreasing trend until it reaches 
around $0.15\text{\AA}$ when MC steps are greater than $70,000$.

Figure~\ref{fgr:bnz}(b) shows errors and RMSD versus MC steps.
Here we set $E_{SS}$  in Eq.~(\ref{eq_relerr}) 
to be $10.60$ kcal/mol as the steady state energy for reference.
The plot shows that 
the errors of total energy are very small
for our MC simulation after $10,000$ iterations.  
When the simulation structure is close to that of its equilibrium state,  the RMSD is smaller than $1\text{\AA}$ and the errors stay in between $1\%$ to $100\%$.
Note since the total energy is a small number, a tiny perturbation causes a large error changing.

Qualitatively, Fig.~\ref{fgr:bnz_vmd}(a) shows that the benzene molecule 
with its initial perturbed structure in blue 
and the equilibrium structure is in green. 
After the MC simulation, we receive the predicted structure in red as compared with the steady state structure in green as shown in 
Fig.~\ref{fgr:bnz_vmd}(b). 
The total CPU time for $100,000$ Monte Carlo steps is $643$ seconds.\\

\subsection{Test case 2: Protein (PDB: 1i2t)}
\begin{figure}
\centering
\includegraphics[height=2.5in]{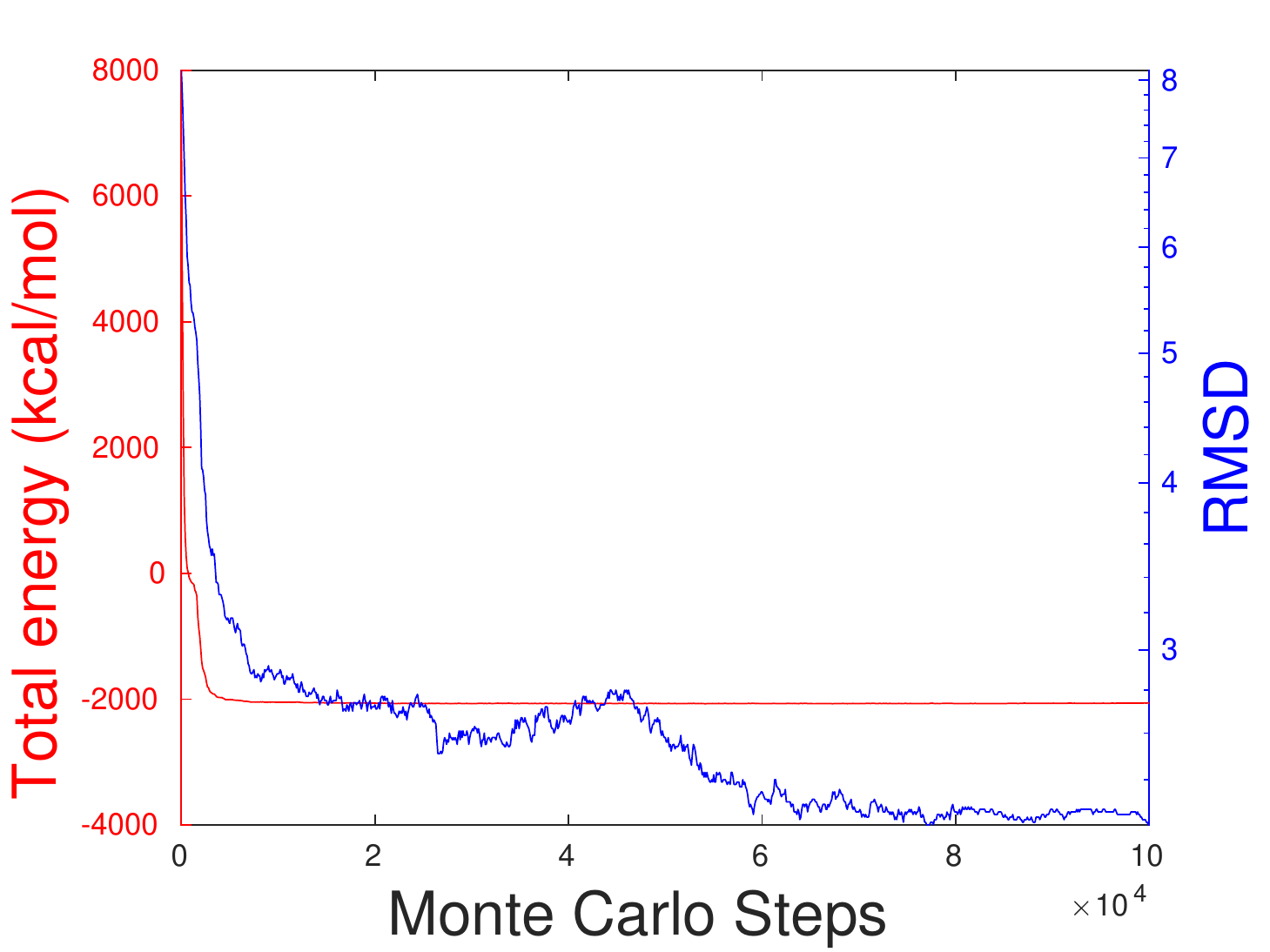}
\includegraphics[height=2.5in]{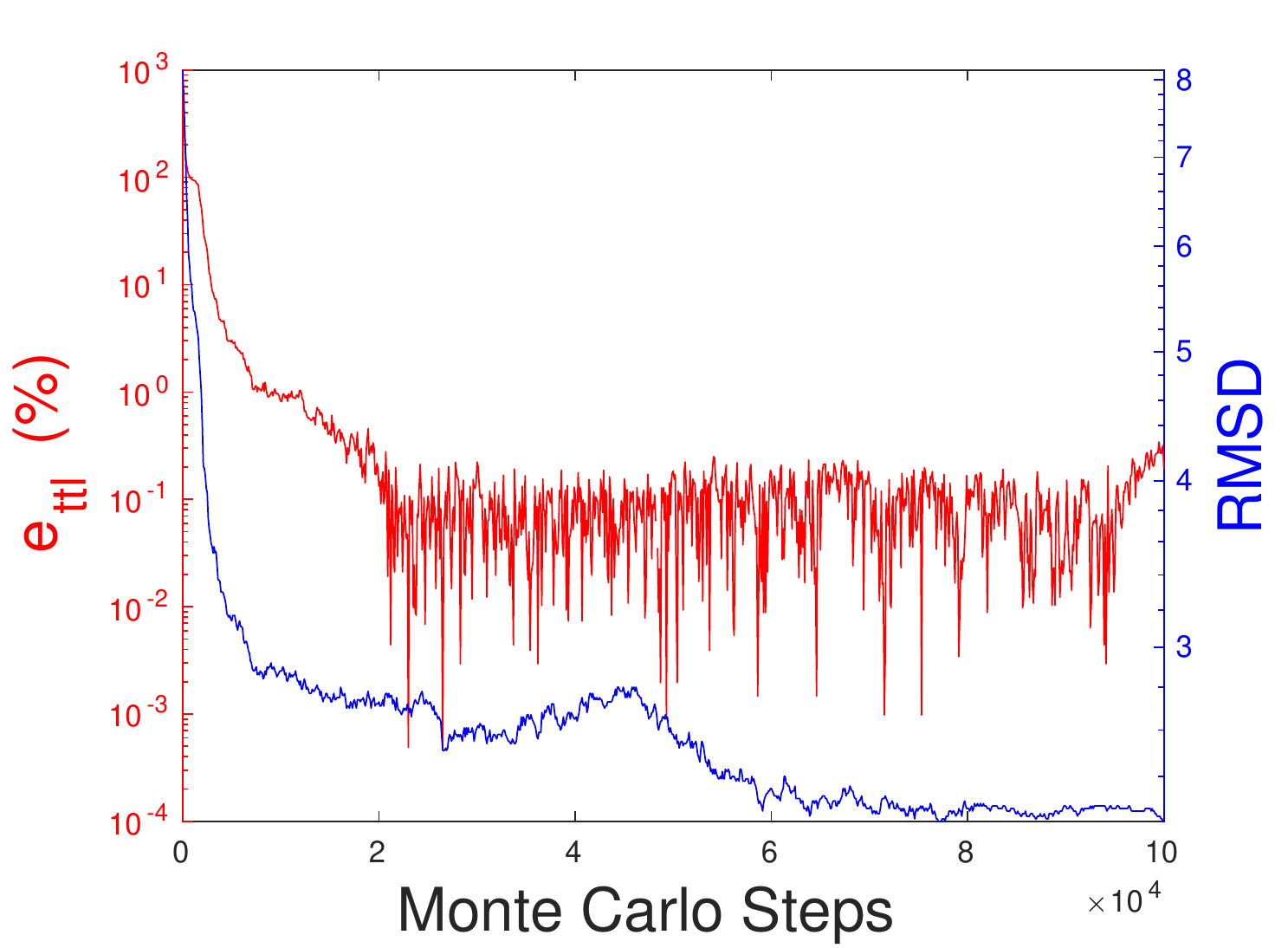}
\begin{picture}(0,0)
\put( -245, 180){(\textbf{a})}
\put( 5, 180){(\textbf{b})}
\end{picture}
\caption{MLIMC  simulation of the protein (PDB: 1i2t) in solvent.
(a) the red curve is the total energy calculated by implicit-solvent Monte Carlo model,
the blue curve is the root mean square deviation of the atomic positions {on each Monte Carlo step to the non-protruded one}; 
(b) the red curve is the error of total energies {e$_\text{ttl}$ defined by Eq.~\ref{eq_relerr}}, 
the blue curve is the same RMSD {as the} left figure.}
\label{fgr:1i2t}
\end{figure}
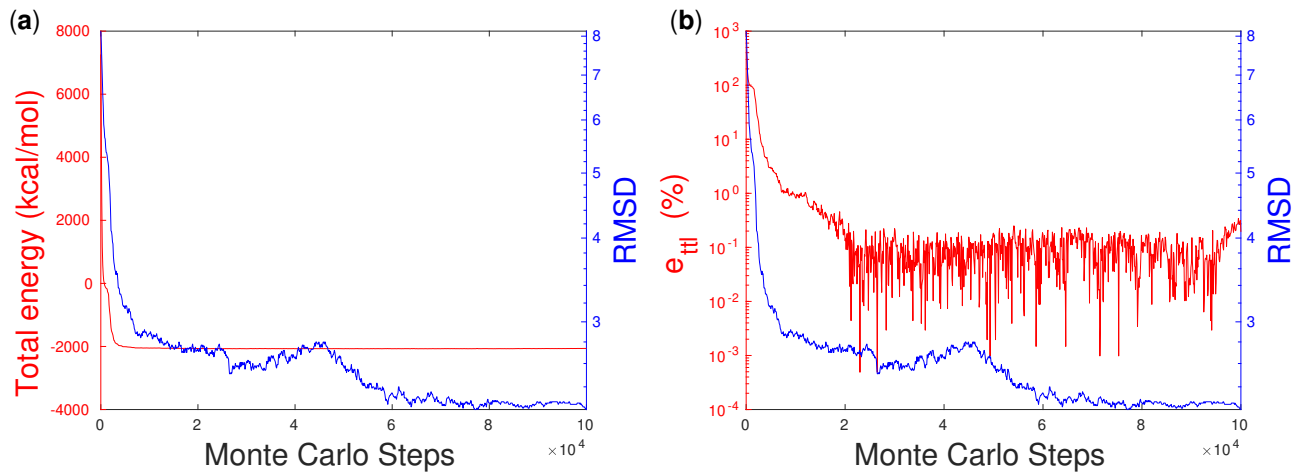

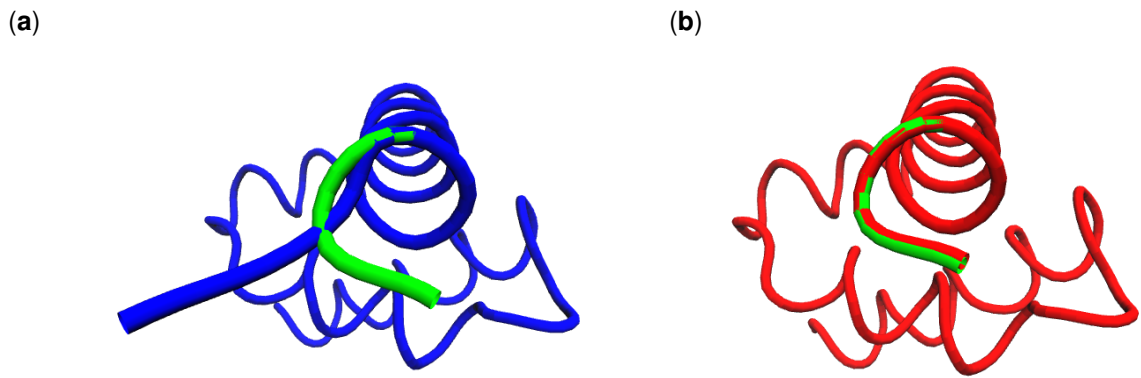
\begin{figure}
\centering
\includegraphics[height=2.6in]{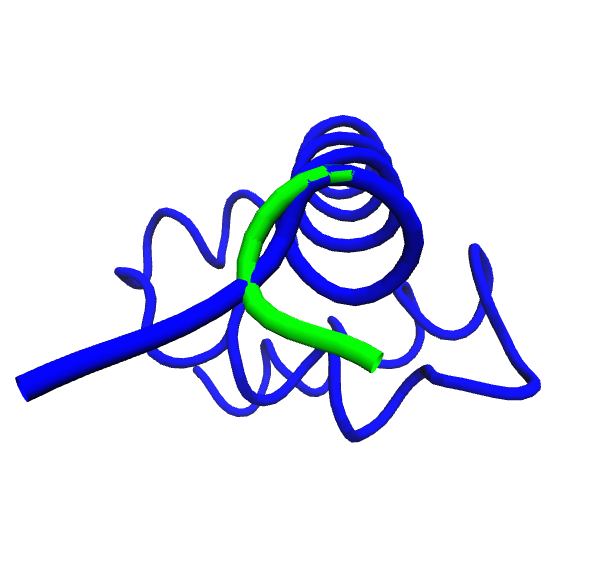}
\includegraphics[height=2.62in]{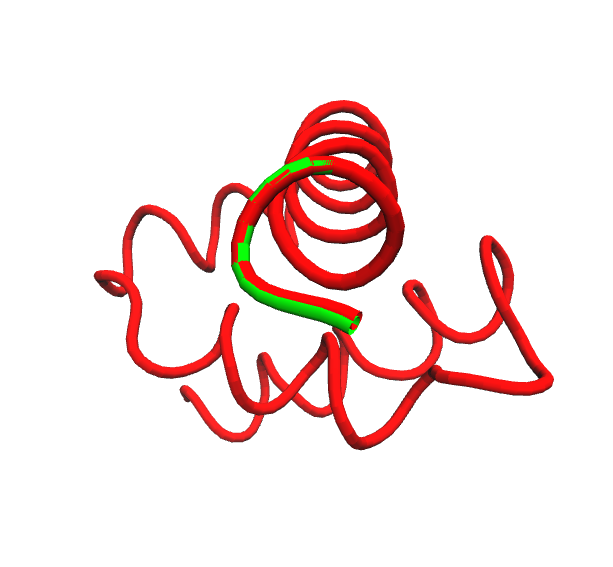}
\begin{picture}(0,0)
\put( -450, 170){(\textbf{a})}
\put( -200, 170){(\textbf{b})}
\end{picture}
\caption{Illustration  of MLIMC simulations of the protein (PDB: 1i2t)~\cite{deo2001x} in solvent.  
(a) the blue structure is the perturbed protein structure generated  with steered molecular dynamics 
and the green one is the original structure at the equilibrium state in solvent.  
(b) the red one is the predicted structure after MLIMC sampling  compared with the original structure in green. Pictures are produced with VMD  \cite{VMD}. }
\label{fgr:1i2tvmd}
\end{figure}

The second MC test is on {the human hyperplastic discs} protein (PDB: 1i2t) with 61 residues. 
We first stretch the last two residues of the original protein 
by a steered molecular dynamics. 
As a result, the stretched molecule has an initial RMSD of $8.14 \text{\AA}$. We apply our MLIMC for  $100,000$ steps, which takes $16,684$ seconds in CPU time. 
Figure~\ref{fgr:1i2t}(a) shows that the total energy of $7260.90$ kcal/mol initially decays rapidly within the first $5,000$ Monte Carlo steps, then oscillates  around $-2070.00$ kcal/mol.
In the same plot, we can see  
the RMSD drops quickly in the first $10,000$ MC steps, after than
decays slowly with fluctuation for the next $40,000$ MC steps, 
then decays steadily after $45,000$ MC steps, 
and finally oscillates slightly around $2.2 \text{\AA}$ after $60,000$ 
MC steps. 
For the energy errors shown in Fig.~\ref{fgr:1i2t} (b), relative to the total energy in the equilibrium of $-2068.13$ kcal/mol, 
the errors rapidly decays in the first $20,000$ MC steps and then 
oscillate within 10\% after that.

Similar to the Benzene case, Fig.~\ref{fgr:1i2tvmd}(a) qualitatively 
shows the perturbed structure in blue against the steady state structure in green for the first two residues and 
Fig.~\ref{fgr:1i2tvmd} (b) shows  that the MLIMC structural prediction in red color after the MC simulation,
which is very close to the steady state structure in green.

\section{Conclusion} \label{sec_conclusion}
 
Monte Carlo simulations are widely used in science and engineering for molecular structure optimization and prediction. 
In many situations, particularly biomolecular systems, the solute molecule is immersed in a water solvent 
and the full-scale explicit solvent Monte Carlo simulations are very expensive. 
Alternatively, implicit solvent Monte Carlo methods using either Poisson-Boltzmann (PB) model or generalized Born (GB) model for computing electrostatics can greatly reduce the degree of freedom.   
However, the accuracy reduction in GB model or the efficiency concerns in PB model hinders the wide application of implicit solvent Monte Carlo simulation.
In this work, we introduce a machine learning-based implicit-solvent Monte Carlo (MLIMC) method for molecular structure optimization and prediction. 
A vital component of our MLIMC is the newly developed Poisson-Boltzmann based machine learning (PBML) model, which maintains the PB accuracy at the GB cost. We validate the proposed MLIMC method by simulating two molecular systems, 
randomly perturbed benzene structure 
and protein (PDB: 1i2t) structures modified by a steered molecular dynamics. 
Numerical experiments demonstrate that proposed MLIMC is efficient in predicting  molecular structures at equilibrium.  
In a comparative analysis, 
we show that the MLIMC model has a great advantage on CPU time and accuracy over DelPhi and Amber PB based Monte Carlo methods. We believe this innovated PBML method can also disruptively change the current status of PB based molecular simulation involving molecular dynamics \cite{geng2011multiscale} and Monte Carlo.  {MLIMC provides accurate electrostatic solvation energy at each configuration of the target protein thus can be helpful in searching protein folding states as intermediate or final using MC based simulation.} The resulting machine learning-based implicit molecular dynamics (MLIMD), together with the present MLIMC model,  will have a vast variety of applications in molecular science, including drug design. 

\vspace*{1cm}
\section*{Acknowledgment}
This work was supported in part by NIH grant  GM126189, NSF grants DMS-2052983,  DMS-1761320, and IIS-1900473,  NASA grant 80NSSC21M0023,  Michigan Economic Development Corporation, MSU Foundation,  Bristol-Myers Squibb 65109, and Pfizer. The work of WG was supported in part by 
NSF grants DMS-1819193 and DMS-2110922.  

\newpage

\section*{Literature cited}
\renewcommand\refname{}

% Use alpha to check for repeated references
%\bibliographystyle{abbrv}
%\bibliographystyle{unsrt}
%\bibliographystyle{ieeetr}
%\bibliography{refs}

\end{document}